\documentclass[11pt,a4paper]{article}
\usepackage[utf8]{inputenc}
\usepackage{amsmath,amssymb,amsfonts}
\usepackage{algorithm}
\usepackage{algorithmic}
\usepackage{graphicx}
\usepackage[margin=1in]{geometry}
\usepackage{setspace}
\usepackage{appendix}
\usepackage{hyperref}

\newcommand{\tr}{\operatorname{tr}}
\newcommand{\SWAP}{\mathrm{SWAP}}

\title{\LARGE{From Betti Numbers to Persistence Diagrams:\\ A Hybrid Quantum Algorithm for Topological Data Analysis}}

\author{Skyler D. Liu\thanks{Corresponding author: liudong@mit.edu}\textsuperscript{\phantom{*}1} \\[0.6em]
{\footnotesize $^{1}$Massachusetts Institute of Technology, Cambridge MA, 02139}}

\date{}

\begin{document}

\maketitle

\begin{abstract}
Persistence diagrams are a central tool in topological data analysis, with applications in pathology, drug discovery, and materials design. Existing quantum algorithms for topological data analysis, most notably the algorithm of Lloyd, Garnerone, and Zanardi (LGZ), efficiently estimate summary statistics such as Betti numbers, but they do not recover persistence diagrams, which record the birth and death scales of individual topological features. This limits their practical use. We propose a hybrid quantum--classical algorithm that predicts persistence diagrams from quantum topological features. The LGZ algorithm serves as a feature extractor: besides Betti numbers, it gives access to quantum states supported on the harmonic subspaces of the combinatorial Laplacians at multiple scales. From these features we construct a positive semidefinite topological kernel and train a quantum least-squares support vector machine (QSVM) that maps the features to a finite set of persistence-diagram classes, whose labels are computed classically on a small training set. Classical persistence computation is thus needed only offline, to label the training set. When persistence in all homology dimensions is required, the prediction cost is polynomial in the number of data points under the assumptions of the LGZ algorithm, which is an exponential speedup over the worst-case cost of exact classical computation.
\end{abstract}

\section{Introduction}

Persistent homology \cite{edelsbrunner2002topological, zomorodian2005computing, edelsbrunner2010computational} is a core method of topological data analysis \cite{carlsson2009topology, ghrist2008barcodes}. It describes the structure of data by tracking how topological features evolve across scales. Its output, the persistence diagram, represents each feature as a point in the plane whose coordinates are the scales at which the feature is born and dies. Persistence diagrams are stable under perturbations of the input \cite{cohen2007stability} and admit a well-developed statistical theory \cite{mileyko2011probability}, which makes them a reliable descriptor in applications such as pathology, drug discovery, and materials design. Their computation, however, is expensive. As discussed by Lloyd et al.\ \cite{lloyd2016quantum}, the best classical algorithms compute the $k$-th Betti number in time $O(n^k)$, and estimating the Betti numbers of all orders to accuracy $\delta$ takes time at least $O(2^n\log(1/\delta))$, where $n$ is the number of vertices. This cost is a major obstacle to applying topological data analysis to large data sets.

Quantum computing \cite{shor1997polynomial}, and quantum machine learning in particular \cite{biamonte2017quantum, schuld2014quest}, offers a possible way around this bottleneck. The LGZ algorithm \cite{lloyd2016quantum} uses quantum phase estimation to estimate Betti numbers of all orders and to diagonalize the combinatorial Laplacian in time polynomial in $n$ (for example, $O(n^5/\delta)$ for the Laplacian), provided that the simplicial complex is sufficiently dense. Compared with the classical $O(2^n\log(1/\delta))$ bound, this suggests an exponential speedup, although the speedup depends on assumptions about the density of the complex and about state preparation. Betti numbers, however, only count topological features; they carry no information about the lifetimes of individual features, which is what most applications require. Ameneyro et al.\ \cite{ameneyro2024quantum} extended the LGZ algorithm to persistent Betti numbers, but this output is still a summary statistic and does not determine the full persistence diagram.

In this paper, we propose an algorithmic pipeline that predicts persistence diagrams from quantum topological features. Our starting point is the observation that the LGZ algorithm gives access to more than Betti numbers. By discrete Hodge theory \cite{eckmann1944harmonische, horak2009spectra}, the zero eigenspace of the combinatorial Laplacian is isomorphic to the real homology group, and every homology class has a unique harmonic representative. Quantum phase estimation in the LGZ algorithm can therefore prepare quantum states supported on these harmonic representatives, which encode where the topological features are located in the complex. We track these states and the Betti numbers across scales, combine them in a positive semidefinite topological kernel, and use the kernel in the quantum support vector machine of Rebentrost et al.\ \cite{rebentrost2014quantum} to predict persistence diagrams.

The pipeline combines classical and quantum computation: classical computation provides labels offline, and quantum computation is used for feature extraction. It targets application scenarios, such as colon lesion detection, in which the relevant topological patterns are finite and can be enumerated. In such settings, a training set that covers the typical patterns can be built. In the training phase, we compute persistence diagrams classically for a limited set of representative data sets and use them as labels, while the LGZ algorithm extracts the quantum topological features. The topological kernel and the quantum least-squares support vector machine then learn the map from the features to the persistence-diagram labels. In the prediction phase, a new data set requires only LGZ feature extraction and kernel-based classification; no persistent homology is computed classically.

Throughout the paper, we consider the setting in which topological features of all homology dimensions $k = 0, 1, \ldots, n-1$ are required. This is the regime in which exact classical computation is exponential in $n$ in the worst case and in which the LGZ algorithm provides an exponential speedup \cite{lloyd2016quantum}. We show that our pipeline retains this speedup at prediction time: its cost is polynomial in $n$, whereas classical persistence computation in all dimensions scales exponentially in the worst case.

\section{Harmonic Feature Extraction}

The first module extracts features from quantum states that carry geometric as well as topological information. Let $D$ be a point cloud of $n$ points. We construct the Vietoris--Rips complexes $X_{\epsilon_1} \subseteq X_{\epsilon_2} \subseteq \cdots \subseteq X_{\epsilon_T}$ at scales $\epsilon_1 < \epsilon_2 < \cdots < \epsilon_T$, where $T$ is the number of scales. At each scale $\epsilon_j$ and for each dimension $k \in \{0, 1, \ldots, K\}$, where $K$ is the maximum homology dimension considered ($K = n-1$ in our main setting, so that all dimensions are covered), we apply the LGZ algorithm \cite{lloyd2016quantum} to the combinatorial Laplacian $\Delta_k$ of $X_{\epsilon_j}$.

Let $S_k$ denote the set of $k$-simplices of the complex and $C_k = \operatorname{span}_{\mathbb{R}}\{|s\rangle : s \in S_k\}$ the corresponding chain space, and let $\partial_k : C_k \to C_{k-1}$ be the boundary operator. The combinatorial Laplacian is
\begin{equation}
\Delta_k = \partial_k^\dagger\partial_k + \partial_{k+1}\partial_{k+1}^\dagger : C_k \to C_k .
\end{equation}
The spectra of such Laplacians reflect the topology of the underlying complex \cite{horak2009spectra, chung1997spectral}. Following the LGZ construction, every $k$-simplex is encoded as an $n$-qubit computational basis state whose $1$-entries mark its $k+1$ vertices. The chain spaces of all complexes in the filtration are therefore subspaces of the same $2^n$-dimensional Hilbert space. The LGZ algorithm works with the Hermitian Dirac operator $B = \sum_k (\partial_k + \partial_k^\dagger)$, whose square is the direct sum of the Laplacians, $B^2 = \bigoplus_k \Delta_k$; hence $B$ and $\Delta_k$ share their null spaces on $C_k$. The algorithm first uses Grover search \cite{grover1996fast} to prepare states supported on $S_k$, in time $O(n^2 \zeta_k^{-1/2})$, where $\zeta_k = |S_k|/\binom{n}{k+1}$ is the fraction of possible $k$-simplices present in the complex. It then applies quantum phase estimation to $e^{-iBt}$. For dense complexes, phase estimation resolves the spectrum of the combinatorial Laplacian to precision $\delta$ in time $O(n^5/\delta)$ \cite{lloyd2016quantum}.

The harmonic subspace $\mathcal{H}_k = \ker \Delta_k$ has a precise topological meaning. By the discrete Hodge decomposition \cite{eckmann1944harmonische, horak2009spectra},
\begin{equation}
C_k = \operatorname{im}\partial_{k+1} \oplus \ker\Delta_k \oplus \operatorname{im}\partial_k^\dagger, \qquad \ker\Delta_k \cong H_k(X;\mathbb{R}),
\end{equation}
so every real homology class has exactly one harmonic representative, and $\beta_k = \dim \ker\Delta_k$ is the $k$-th Betti number. An orthonormal basis of $\mathcal{H}_k$ is mapped to a basis of $H_k(X;\mathbb{R})$. This basis is not canonical, however, and an individual basis vector does not in general correspond to a single geometric hole. Instead, the harmonic subspace as a whole describes where the homology is realized: for $k=1$, harmonic vectors are edge flows that circulate around independent loops, and for $k=2$, they are concentrated on the triangles that enclose cavities.

We therefore use a basis-independent representation of the harmonic subspace. Let $\{|v_i^k(\epsilon_j)\rangle\}_{i=1}^{\beta_k(\epsilon_j)}$ be any orthonormal basis of $\mathcal{H}_k(\epsilon_j)$, and let
\begin{equation}
\Pi_k(\epsilon_j) = \sum_{i=1}^{\beta_k(\epsilon_j)} |v_i^k(\epsilon_j)\rangle\langle v_i^k(\epsilon_j)|
\end{equation}
be the orthogonal projector onto it. Whenever $\beta_k(\epsilon_j) > 0$, we define the harmonic state as the normalized projector
\begin{equation}
\rho_k(\epsilon_j) = \frac{\Pi_k(\epsilon_j)}{\beta_k(\epsilon_j)},
\end{equation}
and we set $\rho_k(\epsilon_j) = 0$ when $\beta_k(\epsilon_j) = 0$. Unlike the uniform superposition $\beta_k^{-1/2}\sum_i |v_i^k\rangle$, which depends on the choice of basis, $\rho_k(\epsilon_j)$ depends only on the subspace $\mathcal{H}_k(\epsilon_j)$. It is also the state that the LGZ procedure produces directly. If phase estimation is applied to the uniform mixture $|S_k|^{-1}\sum_{s\in S_k}|s\rangle\langle s| = I_{C_k}/|S_k|$ and the eigenvalue register is post-selected on the zero eigenvalue, the post-measurement state is
\begin{equation}
\frac{\Pi_k\, (I_{C_k}/|S_k|)\, \Pi_k}{\tr\!\big(\Pi_k\, I_{C_k}/|S_k|\big)} = \frac{\Pi_k}{\beta_k} = \rho_k,
\end{equation}
and the post-selection succeeds with probability $\beta_k/|S_k|$. This identity is exact when the phase-estimation precision is finer than the spectral gap, i.e., when the zero bin contains no nonzero eigenvalue of $B$; the nonzero eigenvalues of $B$ on $C_k$ have magnitude at least $\sqrt{\lambda_{\min}^{+}(\Delta_k)}$, where $\lambda_{\min}^{+}$ denotes the smallest nonzero eigenvalue. Combined with an estimate of $|S_k|$ (for example, by quantum counting), the empirical success frequency also yields an estimate of $\beta_k$. All harmonic states live in the common $n$-qubit Hilbert space, so states of the same data set can be compared directly across scales. For comparisons across data sets, we further assume that the data sets have the same number of points $n$ and that their vertices are indexed consistently (for example, by a common sampling grid or by registration). The harmonic states are not invariant under relabeling of the vertices.

\section{Multi-scale Persistence Features}

Harmonic states at a single scale cannot capture persistence, because persistence describes how features evolve across scales. Motivated by the work of Ameneyro et al.\ \cite{ameneyro2024quantum}, we extract three families of features that describe this evolution from complementary perspectives.

\paragraph{Harmonic state sequences.} For each dimension $k$, we collect the sequence of harmonic states $\rho_k(\epsilon_1), \ldots, \rho_k(\epsilon_T)$. It encodes how the harmonic subspace, and thus the geometric support of the $k$-dimensional features, changes with scale. Between $\epsilon_{j-1}$ and $\epsilon_j$, $\dim\mathcal{H}_k$ changes by the number of $k$-dimensional features born minus the number that die in this interval; each birth adds a new direction to the support of $\rho_k(\epsilon_j)$.

\paragraph{Betti curves.} We record the Betti numbers $\beta_k(\epsilon_j) = \dim\ker\Delta_k(\epsilon_j)$ for every dimension and scale. They count the topological features and complement the harmonic states. Jumps in a Betti curve indicate topological events, and plateaus indicate scale ranges in which the topology is stable. The Betti curve is determined by the persistence diagram, since $\beta_k(\epsilon)$ equals the number of points $(b,d)$ in the dimension-$k$ diagram with $b \le \epsilon < d$. The converse fails: the diagrams $\{(0,2),(1,3)\}$ and $\{(0,3),(1,2)\}$ have the same Betti curve. Betti curves therefore have to be complemented by features that reflect how classes are paired across scales.

\paragraph{Persistence measures.} To quantify the stability of the harmonic subspace between adjacent scales, we define
\begin{equation}
\mathcal{P}_k(j, j+1) = \frac{\tr\!\big(\Pi_k(\epsilon_j)\,\Pi_k(\epsilon_{j+1})\big)}{\sqrt{\beta_k(\epsilon_j)\,\beta_k(\epsilon_{j+1})}} = \frac{\sum_{i} \cos^2\theta_i}{\sqrt{\beta_k(\epsilon_j)\,\beta_k(\epsilon_{j+1})}},
\end{equation}
where $\theta_i$ are the principal angles between $\mathcal{H}_k(\epsilon_j)$ and $\mathcal{H}_k(\epsilon_{j+1})$. We set $\mathcal{P}_k(j,j+1)=1$ if both Betti numbers vanish and $\mathcal{P}_k(j,j+1)=0$ if exactly one of them vanishes. Since there are at most $\min(\beta_k(\epsilon_j),\beta_k(\epsilon_{j+1}))$ principal angles, $0 \le \mathcal{P}_k(j,j+1) \le 1$, with equality to $1$ if and only if the two harmonic subspaces coincide. Because $\tr(\Pi_k(\epsilon_j)\Pi_k(\epsilon_{j+1})) = \beta_k(\epsilon_j)\beta_k(\epsilon_{j+1})\tr(\rho_k(\epsilon_j)\rho_k(\epsilon_{j+1}))$, the measure can be estimated with a swap test on the two harmonic states (Section~\ref{sec:kernel}). A value close to $1$ indicates that the $k$-dimensional structure is stable between the two scales, and a marked decrease signals the birth or death of features. The measure is a heuristic proxy for persistence: even when no feature is born or dies, the harmonic representative of a persisting class generally changes as simplices are added, so $\mathcal{P}_k$ can fall below $1$. Unlike the bottleneck and Wasserstein distances between persistence diagrams \cite{kerber2017geometry}, it is not a metric on diagrams. Its role is to provide the learner with scale-resolved information about where topological events occur.

\paragraph{Cross-dimensional coupling.} Topological events in different dimensions are coupled. In a filtration with field coefficients, each added $k$-simplex either creates a new $k$-dimensional class or destroys a $(k-1)$-dimensional class. Consequently, births in the dimension-$k$ diagram are caused by $k$-simplices and deaths by $(k+1)$-simplices, so features of adjacent dimensions are informative about each other. Features from several dimensions also separate shapes that a single dimension cannot. For example, the torus and the sphere both satisfy $\beta_2 = 1$, so the $H_2$ Betti curve alone may not distinguish them, whereas $\beta_1 = 2$ for the torus and $\beta_1 = 0$ for the sphere. We therefore use features from all dimensions $k \le K$ jointly, which allows the model to exploit cross-dimensional patterns.

\section{Mixed Topological Kernel}
\label{sec:kernel}

The kernel links quantum feature extraction to support vector machine classification and defines the notion of similarity between data sets. Kernels built on persistence diagrams have proved effective in machine learning \cite{kusano2016persistence, reininghaus2015stable}. For two data sets $D_i$ and $D_j$, we define the mixed topological kernel as
\begin{equation}
k_{\text{topo}}(D_i, D_j) = \lambda_1 k_{\text{harm}}(D_i, D_j) + \lambda_2 k_{\text{betti}}(D_i, D_j) + \lambda_3 k_{\text{pers}}(D_i, D_j),
\end{equation}
where the weights $\lambda_1, \lambda_2, \lambda_3 \ge 0$ are selected by cross-validation.

The harmonic kernel compares the harmonic states of two data sets dimension by dimension and scale by scale:
\begin{equation}
k_{\text{harm}}(D_i, D_j) = \sum_{k=0}^{K} \sum_{t=1}^{T} \tr\!\big(\rho_k^{(i)}(\epsilon_t)\,\rho_k^{(j)}(\epsilon_t)\big).
\end{equation}
Each term is a Hilbert--Schmidt inner product between density operators, so $k_{\text{harm}}$ is positive semidefinite. Each term can be estimated with a swap test \cite{buhrman2001quantum}: applied to $\rho\otimes\sigma$, the swap test returns outcome $0$ with probability $\tfrac{1}{2}\big(1+\tr(\rho\sigma)\big)$. A single run acts on $2n+1$ qubits and requires one fresh copy of each state. Terms in which one of the Betti numbers vanishes are zero by definition and need no swap test. Estimating $\tr(\rho\sigma)$ to additive precision $\eta$ requires $O(1/\eta^2)$ repetitions; with amplitude estimation, $O(1/\eta)$ coherent repetitions suffice.

The Betti kernel is the linear kernel on the vectorized Betti curves $\mathbf{v}_{\text{betti}} = \big(\beta_k(\epsilon_t)\big)_{k,t} \in \mathbb{R}^{(K+1)T}$:
\begin{equation}
k_{\text{betti}}(D_i, D_j) = \big\langle \mathbf{v}_{\text{betti}}^{(i)}, \mathbf{v}_{\text{betti}}^{(j)} \big\rangle .
\end{equation}
It captures how the counts of topological features evolve and helps to discriminate between data sets with different topology.

The persistence kernel compares how stable the topology of two data sets is across scales. The persistence measures of each data set are collected in $\mathbf{v}_{\text{pers}} = \big(\mathcal{P}_k(t,t+1)\big)_{k,t} \in [0,1]^{(K+1)(T-1)}$, and we use a Gaussian kernel
\begin{equation}
k_{\text{pers}}(D_i, D_j) = \exp\!\left(-\frac{\big\|\mathbf{v}_{\text{pers}}^{(i)} - \mathbf{v}_{\text{pers}}^{(j)}\big\|_2^2}{2\sigma^2}\right),
\end{equation}
where $\sigma > 0$ is the bandwidth. This kernel responds to small differences in the stability profiles and therefore helps to separate features with different lifetime patterns.

All three kernels are positive semidefinite, and a nonnegative combination of positive semidefinite kernels is positive semidefinite. Hence $k_{\text{topo}}$ is a valid kernel, and every Gram matrix it produces is positive semidefinite.

\section{Quantum Support Vector Machine Training}

The training phase builds a classifier that maps the quantum topological features extracted by the LGZ algorithm to persistence diagrams. We use the quantum least-squares support vector machine (LS-SVM) of Rebentrost et al.\ \cite{rebentrost2014quantum}.

\paragraph{Training set.} The training set $\mathcal{D}_{\text{train}} = \{(D_i, y_i)\}_{i=1}^{M}$ is collected for a specific application scenario, such as colon lesion detection. For each $D_i$, the LGZ algorithm provides the features $\mathbf{X}_i$ (harmonic states, Betti numbers, and persistence measures). The labels are obtained by computing persistence diagrams classically, for example with Ripser \cite{bauer2021ripser}. The diagrams are then grouped into $L$ classes, either by domain-defined categories or by clustering under the bottleneck or Wasserstein distance \cite{kerber2017geometry}, and each class $\ell$ is represented by a prototype diagram $\mathrm{PD}_\ell$ (for example, the medoid of the class). The label $y_i \in \{1, \ldots, L\}$ is the class of the diagram of $D_i$. Alternative vectorizations of persistence diagrams, such as algebraic functions on barcodes \cite{adcock2014ring}, can be used for the clustering step. We assume that, for a fixed scenario, the number of classes $L$ is small, and hence so is the number of classical persistence computations needed to build a training set that covers the typical topological patterns of the scenario.

\paragraph{Gram matrix.} The Gram matrix $\mathbf{G} \in \mathbb{R}^{M\times M}$ has entries $\mathbf{G}_{ij} = k_{\text{topo}}(D_i, D_j)$. Because $k_{\text{topo}}$ is nonlinear in the amplitude-encoded features, $\mathbf{G}$ cannot be obtained as a partial trace of a single data state, as it can for the linear kernel in \cite{rebentrost2014quantum}. Instead, its $M(M+1)/2$ distinct entries are estimated once: the harmonic terms by swap tests, and the Betti and persistence terms classically from the estimated scalar features. The entries are then stored in a quantum random access memory \cite{giovannetti2008quantum}. The LS-SVM below only requires the regularized matrix $\mathbf{A} = \mathbf{G} + \gamma^{-1}I$. The exact $\mathbf{G}$ is positive semidefinite, but the estimated matrix $\tilde{\mathbf{G}} = \mathbf{G} + \mathbf{E}$ need not be. Nevertheless, by Weyl's inequality, $\tilde{\mathbf{A}} = \tilde{\mathbf{G}} + \gamma^{-1}I$ remains positive definite whenever $\|\mathbf{E}\|_2 < \gamma^{-1}$, which holds if every entry is estimated to precision below $\gamma^{-1}/M$. Then $\hat{\mathbf{A}} = \tilde{\mathbf{A}}/\tr(\tilde{\mathbf{A}})$ is a density matrix, and it can be exponentiated by density matrix exponentiation \cite{lloyd2014quantum, rebentrost2014quantum}:
\begin{equation}
\tr_1\!\Big\{e^{-i\,\SWAP\,\Delta t}\,\big(\hat{\mathbf{A}} \otimes \rho\big)\,e^{i\,\SWAP\,\Delta t}\Big\} = e^{-i\hat{\mathbf{A}}\Delta t}\,\rho\, e^{i\hat{\mathbf{A}}\Delta t} + O(\Delta t^2),
\end{equation}
where $\SWAP = \sum_{m,m'=1}^{M} |m\rangle\langle m'| \otimes |m'\rangle\langle m|$ is the swap operator. Each step acts on $O(\log M)$ qubits and is implemented with $O(\log M)$ elementary gates. Simulating $e^{-i\hat{\mathbf{A}}t}$ to error $\varepsilon_{\text{sim}}$ requires $O(t^2/\varepsilon_{\text{sim}})$ steps, each consuming one copy of $\hat{\mathbf{A}}$.

\paragraph{One-vs-rest LS-SVM.} For the multi-class problem, we train $L$ binary classifiers with the one-vs-rest strategy. Classifier $\ell$ solves the linear system
\begin{equation}
\mathbf{F} \begin{pmatrix} b^{(\ell)} \\ \boldsymbol{\alpha}^{(\ell)} \end{pmatrix} = \begin{pmatrix} 0 \\ \mathbf{y}^{(\ell)} \end{pmatrix}, \qquad \mathbf{F} = \begin{pmatrix} 0 & \mathbf{1}^\top \\ \mathbf{1} & \mathbf{A} \end{pmatrix}, \quad \mathbf{A} = \mathbf{G} + \gamma^{-1} I,
\end{equation}
where $\gamma > 0$ is the regularization parameter and $y_i^{(\ell)} = +1$ if $y_i = \ell$ and $y_i^{(\ell)} = -1$ otherwise. The matrix $\mathbf{F}$ is the same for all $L$ classifiers; only the right-hand side changes. Since $\mathbf{A}$ is positive definite, the Schur complement of $\mathbf{A}$ in $\mathbf{F}$, namely $-\mathbf{1}^\top\mathbf{A}^{-1}\mathbf{1}$, is strictly negative. Hence $\det\mathbf{F} = -\det(\mathbf{A})\,\mathbf{1}^\top\mathbf{A}^{-1}\mathbf{1} \neq 0$, so $\mathbf{F}$ is invertible and the system has a unique solution. By Haynsworth's inertia additivity, $\mathbf{F}$ has $M$ positive eigenvalues and exactly one negative eigenvalue.

\paragraph{Quantum linear-system solver.} The system is solved with the HHL algorithm \cite{harrow2009quantum}. Phase estimation with respect to $e^{-i\hat{\mathbf{F}}t}$, where $\hat{\mathbf{F}} = \mathbf{F}/\tr(\mathbf{F})$ (note $\tr\mathbf{F} = \tr\mathbf{A} > 0$) is simulated by combining density matrix exponentiation of $\hat{\mathbf{A}}$ with the remaining sparse terms through the Lie product formula \cite{rebentrost2014quantum}, decomposes the normalized right-hand side $|0,\mathbf{y}^{(\ell)}\rangle$ in the eigenbasis of $\hat{\mathbf{F}}$. A controlled rotation then multiplies each eigencomponent by $\mu_j^{-1}$, where $\mu_j$ are the eigenvalues of $\hat{\mathbf{F}}$; the sign of $\mu_j$ is kept, which handles the negative eigenvalue. Uncomputing the phase register and post-selecting yields the normalized solution state $|b^{(\ell)}, \boldsymbol{\alpha}^{(\ell)}\rangle$. Eigenvalues with $|\mu_j| < \epsilon_{\mathbf{F}}$ are discarded, which defines the effective condition number $\kappa_{\text{eff}} = 1/\epsilon_{\mathbf{F}}$. Following the analysis of Rebentrost et al.\ \cite{rebentrost2014quantum}, and given efficient preparation of $\hat{\mathbf{G}}$ from quantum random access memory, one copy of the solution state for a single classifier is produced in time $O(\log(M)\,\kappa_{\text{eff}}^3/\eta^3)$, where $\eta$ is the target precision. For all $L$ classifiers, the cost is $O(L\log(M)\,\kappa_{\text{eff}}^3/\eta^3)$ per copy. To use the classifiers at prediction time, the $M+1$ coefficients $(b^{(\ell)}, \boldsymbol{\alpha}^{(\ell)})$ are read out to $\ell_2$-precision $\eta$ by state tomography, which requires $\tilde{O}(M/\eta^2)$ copies of each solution state. Because the one-vs-rest rule compares the outputs of different classifiers, the norm of each solution vector is also needed; it is estimated from the post-selection success probability of the HHL algorithm. All of these costs are incurred once, offline.

\section{Prediction}

In the prediction phase, the persistence diagram of a new data set $D_{\text{new}}$ is predicted without any classical persistent homology computation.

\paragraph{Feature extraction.} At each of the $T$ scales and for each dimension $k \le K$, the LGZ algorithm prepares the harmonic state $\rho_k^{\text{new}}(\epsilon_t)$ and estimates $\beta_k^{\text{new}}(\epsilon_t)$. Swap tests between harmonic states at adjacent scales give the persistence measures $\mathcal{P}_k^{\text{new}}(t,t+1)$. We write $C_{\text{LGZ}}$ for the cost of preparing one copy of a harmonic state. For dense complexes, $C_{\text{LGZ}} = O(n^5/\delta)$ \cite{lloyd2016quantum}, up to a post-selection overhead of $O\big((|S_k|/\beta_k)^{1/2}\big)$ when amplitude amplification is used. This overhead is moderate whenever the harmonic subspace is not an exponentially small fraction of the chain space.

\paragraph{Decision function.} For the LS-SVM, the labels enter only through the right-hand side of the linear system, so the coefficients $\boldsymbol{\alpha}^{(\ell)}$ already carry the class signs. The decision function of classifier $\ell$ is therefore
\begin{equation}
f^{(\ell)}(D_{\text{new}}) = b^{(\ell)} + \sum_{i=1}^{M} \alpha_i^{(\ell)}\, k_{\text{topo}}(D_i, D_{\text{new}}).
\end{equation}
The $M$ kernel values $k_{\text{topo}}(D_i, D_{\text{new}})$ are shared by all $L$ classifiers. The harmonic part of each value requires $(K+1)T$ overlaps $\tr\big(\rho_k^{(i)}(\epsilon_t)\rho_k^{\text{new}}(\epsilon_t)\big)$, each estimated with $O(1/\eta^2)$ swap tests. Each swap test consumes one fresh copy of a training harmonic state, which is regenerated from the stored point cloud $D_i$, and one fresh copy of a new-data harmonic state. The Betti and persistence parts are evaluated classically from the stored training features and the estimated features of $D_{\text{new}}$. The predicted class and persistence diagram are
\begin{equation}
\hat{y}_{\text{new}} = \arg\max_{\ell \in \{1, \ldots, L\}} f^{(\ell)}(D_{\text{new}}), \qquad \widehat{\mathrm{PD}}(D_{\text{new}}) = \mathrm{PD}_{\hat{y}_{\text{new}}}.
\end{equation}

\paragraph{Prediction cost.} Kernel evaluation dominates the cost of prediction, and the total cost is
\begin{equation}
O\!\left(\frac{M\,T\,(K+1)\,C_{\text{LGZ}}}{\eta^2} + LM\right) = O\!\left(\frac{M\,T\,(K+1)\,n^5}{\delta\,\eta^2} + LM\right).
\end{equation}
The first term counts the swap tests and the state preparations they require, and the second term counts the classical evaluation of the $L$ decision functions. The cost is polynomial in $n$, $M$, $T$, and $K$, and no classical persistent homology is computed. Algorithm~\ref{alg:predict} summarizes the prediction phase.

\begin{algorithm}[t]
\caption{Prediction of the persistence diagram of a new data set}
\label{alg:predict}
\begin{algorithmic}[1]
\REQUIRE New point cloud $D_{\text{new}}$; training point clouds $\{D_i\}_{i=1}^M$ with stored Betti and persistence features; coefficients $\{b^{(\ell)}, \boldsymbol{\alpha}^{(\ell)}\}_{\ell=1}^{L}$; prototypes $\{\mathrm{PD}_\ell\}_{\ell=1}^{L}$
\FOR{$t = 1, \ldots, T$ and $k = 0, \ldots, K$}
\STATE Estimate $\beta_k^{\text{new}}(\epsilon_t)$ with LGZ phase estimation on $\Delta_k(\epsilon_t)$
\STATE Estimate $\mathcal{P}_k^{\text{new}}(t-1,t)$ with swap tests between $\rho_k^{\text{new}}(\epsilon_{t-1})$ and $\rho_k^{\text{new}}(\epsilon_t)$ (if $t>1$)
\ENDFOR
\FOR{$i = 1, \ldots, M$}
\STATE Estimate $k_{\text{harm}}(D_i, D_{\text{new}})$ with swap tests between $\rho_k^{(i)}(\epsilon_t)$ and $\rho_k^{\text{new}}(\epsilon_t)$ for all $k$ and $t$
\STATE Compute $k_{\text{betti}}(D_i, D_{\text{new}})$, $k_{\text{pers}}(D_i, D_{\text{new}})$, and $k_{\text{topo}}(D_i, D_{\text{new}})$
\ENDFOR
\STATE $\hat{y}_{\text{new}} \leftarrow \arg\max_{\ell} \big(b^{(\ell)} + \sum_{i} \alpha_i^{(\ell)} k_{\text{topo}}(D_i, D_{\text{new}})\big)$
\RETURN $\mathrm{PD}_{\hat{y}_{\text{new}}}$
\end{algorithmic}
\end{algorithm}

\section{Algorithm Analysis}

\subsection{Complexity Analysis}

We compare the complexities for the worst case of a complete simplicial complex on $n$ vertices, which contains all $m = 2^n - 1$ nonempty simplices. This worst case corresponds to our main setting, in which homology in all dimensions ($K = n-1$) is required.

In the classical setting, persistent homology is computed by building the boundary matrix of the filtration and reducing it to obtain the persistence pairs. The standard reduction algorithm takes $O(m^3)$ time, which is $O(2^{3n})$ for the complete complex. Optimizations such as clearing (also called twisting) and cohomology-based reduction \cite{bauer2021ripser} reduce the running time substantially in practice, but they do not improve this worst-case bound. Fast matrix multiplication lowers it to $O(m^\omega) = O(2^{\omega n})$ with $\omega < 2.373$ \cite{milosavljevic2011zigzag}, which is still exponential in $n$. Even without persistence pairs, estimating the Betti numbers of all orders classically requires at least $O(2^n\log(1/\delta))$ time \cite{lloyd2016quantum}.

The proposed algorithm moves the exponential cost into an offline training phase. Training requires classical persistence diagrams for the $M$ training sets, at a one-time worst-case cost of $O(M\cdot 2^{3n})$. Optimized software reduces this cost substantially in practice, and the trained model can be reused. Estimating the Gram matrix and training the QSVM cost $O\big(M^2T(K+1)n^5/(\delta\eta^2)\big)$ and $O\big(LM\log(M)\kappa_{\text{eff}}^3/\eta^5\big)$, respectively, including the readout of the coefficients; both are polynomial. At prediction time, the cost is $O\big(MT(K+1)n^5/(\delta\eta^2) + LM\big)$, which for $K = n-1$ becomes $O\big(MTn^6/(\delta\eta^2) + LM\big)$, where $T$ is the number of scales, $L$ the number of classes, $M$ the number of training sets, and $\delta$ and $\eta$ the precision parameters of the LGZ algorithm and of the swap-test and QSVM estimates.

As an illustration, consider $n = 30$ vertices. For the complete complex, the worst-case cost of standard classical reduction is $2^{3n} = 2^{90} \approx 10^{27}$ operations. On the quantum side, the dominant factor of one LGZ state preparation is $n^5 = 30^5 \approx 2.4\times 10^{7}$, so with $T = 20$ scales and all $K + 1 = n = 30$ dimensions, prediction requires on the order of $M\,T\,n\cdot n^5 \approx 1.5\times 10^{10}\,M$ operations, up to precision-dependent factors. The comparison is therefore between an exponential cost and a polynomial one, which is the exponential speedup that the pipeline inherits from the LGZ algorithm.

This speedup is specific to the all-dimension setting. If only dimensions $k \le K$ for a fixed $K$ are required, classical algorithms can restrict to the $(K+1)$-skeleton, which has $O(n^{K+2})$ simplices, so that standard reduction costs $O(n^{3K+6})$. This cost is polynomial in $n$, and the gain of the quantum pipeline is then polynomial as well. Beyond acceleration, the LGZ algorithm also gives access to harmonic states, which carry geometric information that Betti numbers do not capture and which our method uses to predict persistence diagrams.

\subsection{Applicability}

The method relies on the topological patterns of the target scenario being finite in number. In biomedical applications \cite{nicolau2011topology, bendich2016persistent}, the topological patterns that distinguish normal from diseased samples are often relatively stable. In materials science, the molecular topology of a given class of materials can often be enumerated. In network analysis, domain-specific networks exhibit a limited range of topological features. In such scenarios, the assumptions of the method are plausible, and a training set that covers the relevant patterns can be constructed.

\section{Conclusion}

This work proposes a pipeline that predicts persistence diagrams from quantum topological features, as a step beyond the estimation of summary statistics in quantum topological data analysis. The pipeline combines the harmonic states and Betti numbers obtained from the LGZ algorithm with quantum support vector machine learning \cite{rebentrost2014quantum, biamonte2017quantum}, to map these features to persistence diagram classes.

The central idea is that the harmonic subspaces accessed by the LGZ algorithm carry more information than Betti numbers. Their dimensions are the Betti numbers, and their supports describe where the corresponding homology classes are realized in the complex. By tracking these subspaces across scales and learning from them, we aim to infer persistence diagram patterns while keeping the polynomial cost of quantum feature extraction.

The method is intended for settings in which the topological patterns are limited, such as pathology, drug discovery, and materials design. In these settings, a trained model could be used to screen the topology of a large number of samples.

In practice, the scale grid and the kernel hyperparameters need to be chosen for each application, for example by cross-validation. In addition, the predicted diagram is the prototype of the predicted class, so its resolution depends on how finely the classes are defined for the target scenario.

As quantum hardware matures, the algorithm may become useful for topological analysis and screening in scenarios that involve exponentially many simplices.

\bibliography{references}

\end{document}